\documentclass[sigconf]{acmart}
\usepackage{amsmath,amsfonts}
\usepackage{algorithmic}
\usepackage{graphicx}
\usepackage{textcomp}
\usepackage{xcolor}
\usepackage{graphicx}
\usepackage{listings}
\usepackage{multirow}  
\usepackage{pgfplots}
\pgfplotsset{compat=1.18}
\usepackage{tikz}
\usepackage{booktabs}
\usepackage{hyperref} 
\usepackage{enumitem}
\usepackage{xcolor}
\usepackage[table]{xcolor}

\usepackage{tikz}
\usetikzlibrary{arrows.meta,decorations.pathreplacing,fit,shapes.geometric,calc}
\usepackage{quantikz}

\AtBeginDocument{%
  }

\setcopyright{acmlicensed}
\acmDOI{XXXXXXX.XXXXXXX}

  \acmConference[Q-SE 2026]{7th International Workshop on Quantum Software Engineering}{April 12-18, 2026} {Rio de Janeiro, Brazil}
\acmYear{2026}
\copyrightyear{2026}
\acmISBN{978-1-4503-XXXX-X/2018/06}

\begin{document}

\title{Migrating QAOA from Qiskit 1.x to 2.x: An experience report}

\author{Julien cardinal}
\email{julien.cardinal.1@ens.etsmtl.ca}
\affiliation{%
  \institution{École de Technologie Supérieure de Montréal}
  \city{Montreal}
  \state{Quebec}
  \country{Canada}
}

\author{Imen Benzarti}
\email{imen.benzarti@etsmtl.ca}
\affiliation{%
  \institution{École de Technologie Supérieure de Montréal}
  \city{Montreal}
  \state{Quebec}
  \country{Canada}
}

\author{Ghizlane El boussaidi}
\email{ghizlane.elboussaidi@etsmtl.ca}
\affiliation{%
  \institution{École de Technologie Supérieure de Montréal}
  \city{Montreal}
  \state{Quebec}
  \country{Canada}
}

\author{Christophe Pere}
\email{christophe.pere@etsmtl.ca}
\affiliation{%
 \institution{École de Technologie Supérieure de Montréal}
 \city{Montreal}
  \state{Quebec}
  \country{Canada}
}

\renewcommand{\shortauthors}{Cardinal et al.}

\begin{abstract}
  Migrating quantum algorithms across evolving frameworks introduces subtle behavioral changes that affect accuracy and reproducibility. This paper reports our experience converting the Quantum Approximate Optimization Algorithm (QAOA) from Qiskit Algorithms with Qiskit 1.x (v1 primitives) to a custom implementation using Qiskit 2.x (v2 primitives). Despite identical circuits, optimizers, and Hamiltonians, the new version produced drastically different results. A systematic analysis revealed the root cause: the sampling budget—the number of circuit executions (shots) per iteration. The library’s implicit use of unlimited shots yielded dense probability distributions, whereas the v2 default of $10 000$ shots captured only 23\% of the state space. Increasing shots to $250 000$ restored library-level accuracy. This study highlights how hidden parameters at the quantum–classical interaction level can dominate hybrid algorithm performance and provides actionable recommendations for developers and framework designers to ensure reproducible results in quantum software migration.
\end{abstract}

\ccsdesc[500]{Software and its engineering~Software libraries and repositories}
\ccsdesc[500]{Theory of computation~Probabilistic computation}
\ccsdesc[300]{Theory of computation~Quantum query complexity}
\ccsdesc[100]{Software and its engineering~Software evolution}
\ccsdesc[500]{General and reference~Experimentation}
\ccsdesc[300]{General and reference~Performance}

\keywords{Quantum Optimization, QAOA, Software Architecture Recovery problem, Quantum algorithms}

\received{20 February 2007}
\received[revised]{12 March 2009}
\received[accepted]{5 June 2009}

\maketitle

\section{Introduction}
\label{sec:introduction}
Quantum computing promises significant advantages for solving combinatorial optimization problems that remain intractable for classical computers \cite{maslov2019outlook}. Among the proposed approaches, the Quantum Approximate Optimization Algorithm (QAOA) \cite{farhi2014quantum} is a key hybrid algorithm that alternates between parameterized quantum circuits and classical optimizers to iteratively minimize a cost function.

As quantum software ecosystems evolve, practitioners face a growing challenge: maintaining algorithm implementations across rapidly changing frameworks. High-level quantum libraries, such as Qiskit Algorithms \cite{qiskit_algorithms}, greatly simplify prototyping by abstracting implementation details. However, these abstractions can conceal behaviors that become critical when frameworks evolve. In April 2025, IBM’s Qiskit transitioned from version 1.x to 2.x, deprecating v1 quantum primitives—the low-level execution interfaces that run circuits and collect results—and introducing a new mandatory set of v2 primitives \cite{qiskit_primitives}. This change rendered existing algorithm libraries incompatible with current hardware backends. 

In our ongoing work on quantum optimization for software architecture recovery \cite{belle2015layered}, we encountered this exact migration challenge. To ensure hardware compatibility, we reimplemented QAOA manually using Qiskit 2.x and v2 primitives. What appeared to be a straightforward translation—preserving circuit structure, optimizer settings, and Hamiltonian encoding—revealed subtle but critical differences hidden in high-level abstractions.

Our first custom implementation produced results that deviated drastically from the Qiskit Algorithms baseline: objective values over 100 times higher, no optimal solutions found in ten runs, and complete constraint violations. Extensive verification confirmed that both circuits were unitary-equivalent, used the same optimizer configuration, and encoded the same Hamiltonian. The underlying quantum computation was correct—but the results were not.

A systematic investigation uncovered the true cause: the sampling budget, defined as the number of circuit executions (shots) per optimization iteration. The Qiskit Algorithms library, using a statevector simulator, implicitly evaluates an unlimited number of shots, producing dense probability distributions that cover about 97\% of the solution space. In contrast, our implementation’s default of $10 000$ shots generated sparse distributions covering only 23\% of possible states. 

This paper presents a case study on the migration of QAOA from Qiskit 1.x to 2.x, illustrating how hidden quantum–classical interface parameters can profoundly affect hybrid algorithm performance. We show empirically that accuracy scales monotonically with the sampling budget: 10 000 shots yield 40\% success, $100 000$ shots reach 80\%, and 250 000 shots achieve 97\%, nearly matching the library baseline. We also derive practical migration guidelines for developers and propose framework-level recommendations to improve transparency and reproducibility across future Qiskit releases.
Section~\ref{sec:background} provides background on QAOA, the software architecture recovery problem, and Qiskit's architectural evolution. Section~\ref{sec:Migration-process-and-initial-results} describes our migration process and initial results. 
Section~\ref{sec:systematic-investigation-of-the-root-cause} presents the systematic comparison isolating sampling budget as the root cause. Section~\ref{sec:analysis-and-implications} analyzes implications for hybrid algorithm design and provides actionable recommendations. 
Section~\ref{sec:limitations} discusses limitations and future work.
Section~\ref{sec:related} surveys related work. 
 Section~\ref{sec:conclusion} concludes.

\section{Background and Motivation}
\label{sec:background}

\subsection{Software Architecture Recovery Problem}

The optimization problem considered in this work is the \textit{Layered Architecture Recovery} (LAR) problem \cite{belle2015layered}. The objective is to assign each software package to one architectural layer minimizing undesirable dependencies. In well-structured architectures, dependencies should occur between adjacent layers, while intra-layer, back-layer, and skip-layer dependencies are penalized.

\begin{figure}[htbp]
\centerline{\includegraphics[scale=0.34]{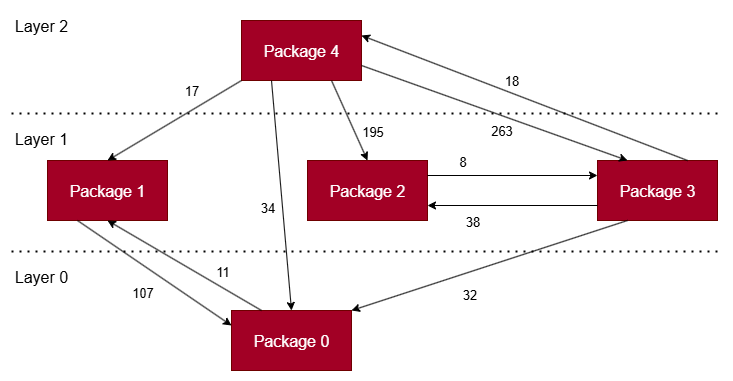}}
\caption{An example of a layered architecture.}
\label{fig:LAR_example}
\end{figure}

Figure~\ref{fig:LAR_example} shows an example of the  architecture of a small utility software. Packages are linked by weighted arrows indicating dependency count and direction. Calls to adjacent lower layers incur no penalty; all other cases are penalized. Example of penalties \cite{belle2015layered}: 1× for skip-calls (calling two or more layers down), 15× for back-calls (calling layers above), and 2× for intra-dependencies (same layer).

This example contains one skip-call (34 dependencies), two back-calls totaling $(11+18)\times15=435$, and two intra-dependencies totaling $(8+38)\times2=92$, yielding a cost of $34+435+92=561$ for this assignment.

The LAR problem is formulated as a Quadratic Semi-Assignment Problem (QSAP) \cite{pardalos1994quadratic} using three matrices: weight matrix $W$ (dependencies between packages), cost matrix $C$ (penalty costs based on layer placement), and binary assignment matrix $X$ (rows=packages, columns=layers). The QSAP matrices for the above example are:

\[
W=
\begin{bmatrix}
0 & 11 & 0 & 0 & 0\\
107 & 0 & 0 & 0 & 0\\
0 & 0 & 0 & 8 & 0\\
32 & 0 & 38 & 0 & 18\\
34 & 17 & 195 & 263 & 0
\end{bmatrix}
\quad
C=
\begin{bmatrix}
2 & 15 & 15\\
0 & 2 & 15\\
1 & 0 & 2
\end{bmatrix}
\]
\[
X=
\begin{bmatrix}
1 & 0 & 0\\
0 & 1 & 0\\
0 & 1 & 0\\
0 & 1 & 0\\
0 & 0 & 1
\end{bmatrix}
\]

From this QSAP form, the problem becomes a \textit{Quadratic Unconstrained Binary Optimization} (QUBO) problem, where binary matrix $X$ becomes vector $x$ and each variable $x_{i\times n+k}$ indicates whether package $i$ is assigned to layer $k$ ($n$ = number of layers). The objective function is:

\begin{equation}
\text{minimize: } \mathbf{x}^T Q \mathbf{x}
\end{equation}

where $Q$ is the QUBO matrix sized $mn \times mn$ ($m$ = packages, $n$ = layers). For the example above, the QUBO matrix is $15 \times 15$. Minimizing this function finds the optimal assignment—the vector $x$ that yields the lowest cost.

Matrix $Q$ combines dependency weights from $W$ with penalty coefficients from $C$, creating a quadratic form suitable for quantum optimization. Once expressed as QUBO, the problem maps to an Ising Hamiltonian—a Hamiltonian containing only one- and two-body spin,  interactions \cite{lucas2014ising}—taking the form:

\begin{equation}
H = \sum_{i} h_i Z_i + \sum_{i<j} J_{ij} Z_i Z_j
\end{equation}

where $Z_i$ are Pauli-Z operators on qubit $i$. Unlike general Hamiltonians that may include arbitrary multi-qubit terms, this restricted form enables efficient quantum algorithm implementation. Figure~\ref{fig:qaoa_flow} shows the transformation pipeline. The LAR problem is formulated as QUBO, translated into an Ising Hamiltonian <H> whose ground state represents the optimal architecture, while QAOA uses a quantum circuit to approximate this state and a classical optimizer adjusts parameters to improve accuracy.

\definecolor{stageA}{RGB}{140,109,179}  
\definecolor{stageB}{RGB}{241,163,64}   
\definecolor{stageC}{RGB}{255,204,92}   
\definecolor{stageD}{RGB}{102,194,165}  
\definecolor{notegray}{RGB}{90,90,90}

\begin{figure*}[t]
\centering
\resizebox{\textwidth}{!}{
\begin{tikzpicture}[
  font=\small,
  box/.style={rounded corners=2mm, draw=black, thick, align=left,
              minimum width=3.0cm, minimum height=1.6cm, inner sep=2.5mm},
  slimbox/.style={rounded corners=2mm, draw=black, thick, align=left,
              minimum width=2.8cm, minimum height=1.3cm, inner sep=2mm},
  arr/.style={-{Latex[length=2mm]}, thick},
  note/.style={align=left, text=notegray, font=\small}
]

\node[box, fill=stageA!25] (prob) {
  \textbf{Layered Architecture}\\
  \textbf{Recovery (LAR)}\\
   Dependency Graph\\
   Goal: assign packages to layers\\
   Penalize back/skip/intra deps\\
};

\node[box, fill=stageB!25, right=0.8cm of prob] (qubo) {
  \textbf{QUBO Formulation}\\
   Binary vars $x_{ik}\!\in\!\{0,1\}$\\
  Minimize $x^\top Q x$
};

\node[box, fill=stageC!35, right=0.8cm of qubo] (ham) {
  \textbf{Ising Hamiltonian}\\
   $H = \sum h_i Z_i + \sum J_{ij} Z_i Z_j$\\
  Encodes QUBO as \\
  quantum energy
};

\node[box, fill=stageD!35, right=0.8cm of ham] (qaoa) {
  \textbf{QAOA Hybrid Optimization}\\
   Alternating $U_C(\gamma), U_M(\beta)$\\
   Measure \& return bitstring samples
};

\node[slimbox, fill=gray!12, above right=0cm and 0.5cm of qaoa.north east] (post) {
    \textbf{Post-processing}\\
   Select top-$k$ bitstrings \\
   by frequency\\
   Evaluate cost \& feasibility
};

\node[slimbox, fill=gray!15, below=0.5cm of qaoa] (opt) {
      \textbf{Classical Optimizer}\\
   Evaluate objective $\langle H_c\rangle$\\
   Update parameters $(\gamma,\beta)$
};

\draw[arr] (prob.east) -- node[above, note]{
} (qubo.west);

\draw[arr] (qubo.east) -- node[above, note]{ 
} (ham.west);

\draw[arr] (ham.east) -- node[above, note]{ 
} (qaoa.west);

\draw[arr] (qaoa.north) |- node[above right, note]{bitstring samples} (post.west);

\draw[arr] (post.south) |- node[right=0.1cm, note]{filtered results} (opt.east);

\draw[arr] (opt.north) -- node[right, note]{ 
} (qaoa.south);

\end{tikzpicture}
}
\caption{Conceptual flow of the QAOA-based hybrid process with explicit \textit{Post-processing}. The Layered Architecture Recovery (LAR) problem is transformed into a QUBO, mapped to an Ising Hamiltonian, and optimized with QAOA. After quantum execution, bitstring samples are classically post-processed to select feasible and low-cost solutions before the optimizer updates $(\gamma,\beta)$ parameters.}
\label{fig:qaoa_flow}
\end{figure*}
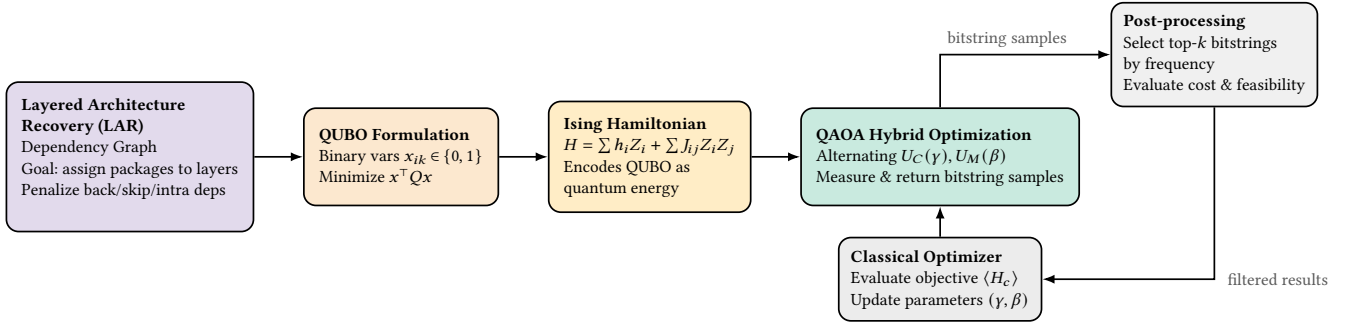

\subsection{Quantum Optimization and the QAOA Approach}

\begin{figure}[t]
\centering
\resizebox{\columnwidth}{!}{%
\begin{quantikz}[row sep=0.35cm, column sep=0.5cm]
\lstick{$q_0$}     & \gate{H} & \gate{U_C(\gamma_1)} \gategroup[4,steps=2,style={dashed,rounded corners,fill=blue!10,inner xsep=2pt},background,label style={label position=below,anchor=north,yshift=-0.2cm}]{{Repetition 1}} & \gate{U_M(\beta_1)} & \gate{U_C(\gamma_2)} \gategroup[4,steps=2,style={dashed,rounded corners,fill=blue!10,inner xsep=2pt},background,label style={label position=below,anchor=north,yshift=-0.2cm}]{{Repetition 2}} & \gate{U_M(\beta_2)} & \meter{} & \rstick[4]{\parbox{3cm}{\centering Classical optimizer updates $(\gamma,\beta)$}}\\
\lstick{$q_1$}     & \gate{H} & \gate{U_C(\gamma_1)} & \gate{U_M(\beta_1)} & \gate{U_C(\gamma_2)} & \gate{U_M(\beta_2)} & \meter{} & \\
\lstick{$\vdots$}  & \vdots   & \vdots                & \vdots               & \vdots               & \vdots               & \vdots   & \\
\lstick{$q_{n-1}$} & \gate{H} & \gate{U_C(\gamma_1)} & \gate{U_M(\beta_1)} & \gate{U_C(\gamma_2)} & \gate{U_M(\beta_2)} & \meter{} & 
\end{quantikz}%
}
\caption{QAOA circuit with \(p=2\) repetitions. The circuit prepares the superposition state \(|+\rangle^{\otimes n}\) with Hadamard gates, alternates problem unitaries \(U_C(\gamma_\ell)\) and mixer unitaries \(U_M(\beta_\ell)\), and measures all qubits. A classical optimizer iteratively updates \((\gamma,\beta)\) to minimize the expected cost.}
\label{fig:qaoa_circuit}
\end{figure}
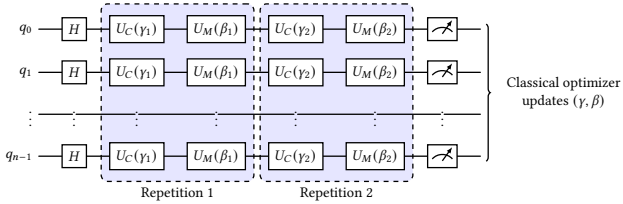

Among the different families of quantum algorithms, the \textit{Quantum Approximate Optimization Algorithm} (QAOA) \cite{farhi2014quantum} is one of the most studied hybrid approaches. QAOA belongs to the class of \textit{Variational Quantum Algorithms} (VQAs)~\cite{cerezo2021variational}, which combine a parameterized quantum circuit with a classical optimizer to minimize a cost function derived from the problem’s Hamiltonian.
As shown in figure~\ref{fig:qaoa_circuit}, the algorithm alternates between two Hamiltonians: a \textit{cost Hamiltonian $H_c$}, which encodes the optimization objective, and a \textit{mixer Hamiltonian $H_m$}, which ensures exploration of the solution space. The parameters \((\gamma,\beta)\) (rotation angles) of these operations are adjusted by a classical optimizer to find a state that minimizes the expected value of the cost operator. This hybrid process makes QAOA a key algorithm for exploring quantum–classical interactions in the current \textit{Noisy Intermediate-Scale Quantum (NISQ)} era~\cite{preskill2018quantum}.

The circuit applies $p$ repetitions of these operators with tunable parameters $(\gamma_\ell, \beta_\ell)$. The quantum state evolves as:

\begin{equation}
\ket{\psi(\boldsymbol{\gamma}, \boldsymbol{\beta})} = \prod_{\ell=1}^{p} e^{-i\beta_\ell H_M} e^{-i\gamma_\ell H_C} \ket{+}^{\otimes n}
\end{equation}

where $\ket{+}^{\otimes n}$ is the uniform superposition initialized by Hadamard gates. A classical optimizer iteratively updates parameters to minimize $\langle H_C \rangle$, the expected energy. After optimization converges, the circuit is measured repeatedly (shots), producing a probability distribution over bitstring solutions. Post-processing evaluates these candidates classically to select the best x.

\subsection{Qiskit Ecosystem Evolution}

IBM’s open-source framework \textit{Qiskit} \cite{javadi2024qiskit} provides an extensive environment for building, simulating, and executing quantum circuits. Over time, the ecosystem has evolved through multiple modules, including:
\begin{itemize}[leftmargin=1.1em, itemsep=0pt, topsep=1pt]
    \item \textbf{Qiskit SDK}: Core circuit manipulation and simulation
    \item \textbf{Qiskit Algorithms}, offering ready-to-use implementations of popular algorithms such as QAOA and VQE;
    \item \textbf{Qiskit Optimization}, providing modeling tools for quantum optimization problems;
    \item \textbf{Qiskit Runtime}, allowing hybrid workflows to run efficiently on IBM Quantum hardware.
\end{itemize}

Recent developments in Qiskit introduced a major architectural change with version 2 (Qiskit 2.x), which replaces legacy execution interfaces requiring the new generation of \textit{quantum primitives} \cite{qiskit_primitives}.  
Primitives define a consistent and hardware-agnostic API for evaluating quantum circuits: (1) \textbf{Estimator} primitive computes expected values of observables for a given parameterized circuit; (2) The \textbf{Sampler} primitive executes a circuit multiple times and returns measurement outcome distributions.


V2 primitives offer unified interfaces for simulators and hardware, explicit configuration (shots, precision, error mitigation), and hardware-optimized execution strategies. However, Qiskit Algorithms was relying on v1 primitives before August 29th 2025, creating incompatibility with current hardware runtime. 


\subsection{Motivation for the custom implementation}

Two main motivations guided the development of a custom implementation. 
First, \textbf{hardware compatibility}: executing the algorithm on IBM quantum processors requires Qiskit~2.x and its new quantum primitives. 
However, the Qiskit Algorithms library version~0.3.1 used in our initial experiments relies on v1 primitives, making it incompatible with current IBM hardware. 
Second, \textbf{algorithmic transparency}: high-level libraries often abstract internal mechanisms such as parameter initialization, circuit optimization, and post-processing strategies. 
A custom implementation provides full control over these components, including sampling and post-processing, while offering visibility into optimization trajectories and a foundation for future algorithmic enhancements.

The migration appeared straightforward: replicate the circuit structure, use equivalent optimizer settings, apply the same Hamiltonian. As we discovered, this surface-level equivalence missed critical differences in the quantum-classical interface.

\section{Migration Process and Initial Results}
\label{sec:Migration-process-and-initial-results}

\subsection{Implementation Approach}

The custom manual QAOA implementation replicates the Qiskit Algorithms library behavior while ensuring compatibility with the latest Qiskit primitives. Both implementations follow the identical workflow shown in Figure~\ref{fig:qaoa_flow}, with only the encoding step reused from the baseline implementation.

\paragraph{Problem Encoding}
The process begins by transforming the optimization problem into a Hamiltonian that can be integrated into a quantum circuit. The input is a QUBO matrix, which is converted into an equivalent Ising Hamiltonian. These formulations describe the same problem but differ in numerical representation. The Hamiltonian form allows direct translation into parameterized quantum operations, enabling implementation using the QAOA circuit. The transformation can be seen in Listing~\ref{lst:qubo_to_ising}

\begin{lstlisting}[language=Python, caption={QUBO to Ising hamiltonian conversion},label={lst:qubo_to_ising}]
# Simplified code
# Create the QUBO
qubo = create_qubo(flow_matrix, distance_matrix)
# Create a quadratic program from a QUBO
quadratic_program = create_quadratic_program(qubo)
# Convert the quadratic profrom to an Ising Hamiltonian
cost_hamiltonian = quadratic_program.to_ising()
\end{lstlisting}

\paragraph{Quantum Circuit Construction}
The quantum part of the algorithm is implemented following the standard QAOA ansatz, shown in Figure~\ref{fig:qaoa_circuit}. The circuit requires as many qubits as the size of the QUBO matrix. A layer of Hadamard gates initializes each qubit in superposition. Then, a fixed number of layers (\(p=5\) in our experiments) alternates between two parameterized Hamiltonians:
\begin{itemize}[leftmargin=1.1em, itemsep=0pt, topsep=1pt]
  \item the \textit{cost Hamiltonian} \(U_C(\gamma)\), which encodes the cost function;
  \item the \textit{mixer Hamiltonian} \(U_M(\beta)\), which explores the search space by rotating qubit states.
\end{itemize}
For this step, the Qiskit \textit{QAOAAnsatz()} function can be used as shown is Listing~\ref{lst:ansatz}


\begin{lstlisting}[language=Python, caption={Creation of the quantum ansatz},label={lst:ansatz}]
from qiskit.circuit.library import QAOAAnsatz
circuit = QAOAAnsatz(cost_hamiltonian, p=5)
# Produces alternating U_C and U_M structure
\end{lstlisting}

\paragraph{Hybrid Quantum--Classical Optimization}
The quantum circuit is evaluated iteratively using Qiskit’s \textit{estimator} primitive, which computes the expected value of the Hamiltonian for a given set of parameters. This expectation value represents the circuit’s energy and is minimized by a classical optimizer. The classical component (COBYLA in this case) updates the parameters \((\gamma, \beta)\) after each quantum evaluation, forming a hybrid feedback loop as illustrated in Figure~\ref{fig:qaoa_flow}.

This step is implemented as seen in Listing~\ref{lst:optimization}. The parameters \((\gamma,\beta)\) are initialized randomly in the range \([-2\pi,2\pi]\), ensuring broad coverage of the parameter space.

\begin{lstlisting}[language=Python, caption={QAOA optimization loop with configurable shots},label={lst:optimization}]
# simplified code 
from qiskit.primitives import Estimator
from scipy.optimize import minimize
import numpy as np

estimator = Estimator()
def cost_function(params, circuit, cost_hamiltonian):
    job = estimator.run([(circuit, cost_hamiltonian, params)])
    return job.result()[0].data.evs
# Random parameter initialization 
init_params = np.random.uniform(-2*np.pi, 2*np.pi, 2*p)
# Optimize
result = minimize(cost_function, init_params, 
                  args=(circuit, cost_hamiltonian),
                  method="COBYLA", 
                  options={"maxiter": 1000})
\end{lstlisting}

\paragraph{Measurement and Post-Processing}
Once the optimizer converges to the lowest observed energy, the final parameters are applied to the circuit. All qubits are then measured, and the circuit is executed 10\,000 times to obtain a probability distribution of bitstring results. Listing~\ref{lst:sampling} shows the operations necessary for this step.  
A post-processing phase interprets these samples by:
\begin{enumerate}[label=\textbf{\arabic*.}, leftmargin=1.2em, itemsep=0pt, topsep=1pt]
  \item aggregating and filtering the most frequent bitstrings to increase accuracy;
  \item translating the bitstrings into assignments for the original optimization problem.
\end{enumerate}
This step is critical, as it determines the final solution quality and will later be shown to account for most of the observed difference in accuracy between the two implementations.

\begin{lstlisting}[language=Python, caption={Post-processing: sampling and ranking solutions},label={lst:sampling}]
# simplified code 
from qiskit.primitives import Sampler

sampler = Sampler()
final_circuit = circuit.assign_parameters(result.x)
final_circuit.measure_all()
# Sample circuit and get probability distribution
job = sampler.run([final_circuit], shots=10000)
counts = job.result()[0].data.meas.get_int_counts()
probabilities = {k: v/sum(counts.values()) for k, v in counts.items()}
# Rank solutions by probability
solutions = sorted(probabilities.items(), key=lambda x: x[1], reverse=True)
\end{lstlisting}

\subsection{Experimental Setup}

For our experiments, we use a $5\times3$ LAR instance (5 packages, 3 layers) requiring 15 qubits with known optimal objective value of 561. Table~\ref{tab:exp_setup} summarizes our experimental configuration.


\begin{table}[htbp]
\caption{Experimental Configuration}
\label{tab:exp_setup}
\centering
\small
\begin{tabular}{llp{4cm}}
\toprule
\textbf{Category} & \textbf{Parameter} & \textbf{Value} \\
\midrule
\multirow{3}{*}{Problem} & LAR instance & 5 packages, 3 layers \\
& QUBO dimension & $15 \times 15$ \\
& Optimal objective & 561 \\
\midrule
\multirow{4}{*}{Algorithm} & QAOA repetitions ($p$) & 5 \\
& Optimizer & COBYLA \\
& Max iterations & 1\,000 \\
& Parameter init & Uniform$[-2\pi, 2\pi]$ \\
\midrule
\multirow{3}{*}{Execution} & Simulator & StatevectorSimulator \\
& Shots (initial) & 10\,000 \\
\bottomrule
\end{tabular}
\end{table}

\subsection{Initial Results}

\paragraph{Library Baseline}
Column \textit{Library Baseline} of Table~\ref{tab:test_library_and_manual} summarizes the performance of the library version. The algorithm consistently found optimal assignments that satisfy all QSAP constraints, with stable objective values and average runtimes near 174.5~seconds.

\begin{table*}[htbp]
\caption{Qiskit Algorithms Performance of 10 Executions of QAOA on the Library Baseline, our Initial Custom Implementation and our Custom Implementation After Post-Processing Threshold Correction}
\centering
\small  
\setlength{\tabcolsep}{3pt} 
\begin{tabular}{c|cccc|cccc|cccc}
\toprule
& \multicolumn{4}{c|}{\textbf{Library Baseline}} & \multicolumn{4}{c|}{\textbf{Initial Custom}} & \multicolumn{4}{c}{\textbf{Custom-Post-Proc-threshold-corrected}} \\
\midrule
\textbf{Run} & \textbf{Shots} & \textbf{Time (s)} & \textbf{Object.} & \textbf{States Ev.} & \textbf{Shots} & \textbf{Time (s)} & \textbf{Object.} & \textbf{States Ev.} & \textbf{Shots} & \textbf{Time (s)} & \textbf{Object.} & \textbf{States Ev.} \\
\midrule
1 & $\infty$ & 196.2 & \cellcolor{green!80!yellow!50}561 & 31420 (96\%) & 10k & 53.6 & 87275 & 10 (0.03\%) & 10k & 61.4 & 649 & 7530 (23\%) \\
2 & $\infty$ & 177.1 & \cellcolor{green!80!yellow!50}561 & 31685 (97\%) & 10k & 57.0 & 48653 & 10 (0.03\%) & 10k & 62.3 & 848 & 7592 (23\%) \\
3 & $\infty$ & 171.8 & \cellcolor{green!80!yellow!50}561 & 31686 (97\%) & 10k & 55.2 & 48929 & 10 (0.03\%) & 10k & 61.6 & \cellcolor{green!80!yellow!50}561 & 7694 (23\%) \\
4 & $\infty$ & 191.2 & \cellcolor{green!80!yellow!50}561 & 31705 (97\%) & 10k & 62.5 & 80792 & 10 (0.03\%) & 10k & 62.5 & \cellcolor{green!80!yellow!50}561 & 7667 (23\%) \\
5 & $\infty$ & 193.2 & \cellcolor{green!80!yellow!50}561 & 29877 (91\%) & 10k & 52.7 & 46336 & 10 (0.03\%) & 10k & 57.4 & 702 & 7602 (23\%) \\
6 & $\infty$ & 168.4 & \cellcolor{green!80!yellow!50}561 & 31130 (95\%) & 10k & 63.0 & 39737 & 10 (0.03\%) & 10k & 67.7 & 649 & 7599 (23\%) \\
7 & $\infty$ & 160.0 & \cellcolor{green!80!yellow!50}561 & 31694 (97\%) & 10k & 64.0 & 83419 & 10 (0.03\%) & 10k & 65.3 & 625 & 7605 (23\%) \\
8 & $\infty$ & 161.6 & \cellcolor{green!80!yellow!50}561 & 31716 (97\%) & 10k & 66.3 & 86528 & 10 (0.03\%) & 10k & 60.1 & 881 & 7295 (22\%) \\
9 & $\infty$ & 157.9 & \cellcolor{green!80!yellow!50}561 & 31683 (97\%) & 10k & 64.0 & 122033 & 10 (0.03\%) & 10k & 64.3 & \cellcolor{green!80!yellow!50}561 & 7512 (23\%) \\
10 & $\infty$ & 167.6 & \cellcolor{green!80!yellow!50}561 & 31742 (97\%) & 10k & 61.8 & 45293 & 10 (0.03\%) & 10k & 59.6 & \cellcolor{green!80!yellow!50}561 & 7651 (23\%) \\
\midrule
\textbf{Mean} & -- & \textbf{174.5} & \textbf{561} & \textbf{31434 (96\%)} & \textbf{10k} & \textbf{60.0} & \textbf{68900} & \textbf{10 (0.03\%)} & \textbf{10k} & \textbf{62.2} & \textbf{660} & \textbf{7575 (23\%)} \\
\textbf{Std Dev} & -- & \textbf{14.3} & \textbf{0} & \textbf{579} & -- & \textbf{4.9} & \textbf{26976} & \textbf{0} & -- & \textbf{3.0} & \textbf{119} & \textbf{113} \\
\midrule
\textbf{QSAP Constr.} & \multicolumn{4}{c|}{\textbf{Constraints satisfied by all runs}} & \multicolumn{4}{c|}{\textbf{None satisfy the constraints}} & \multicolumn{4}{c}{\textbf{Constraints satisfied by all runs}} \\
\bottomrule
\multicolumn{13}{l}{Green cells mean that the objective function results in the optimal solution.}\\
\multicolumn{13}{l}{Run: execution number, Shots: sampling budget, Time (s): execution duration in seconds, Object.: objective function value,}\\
\multicolumn{13}{l}{State Ev.: number of states evaluated classically, QSAP Constr.: QSAP constraints compliance by QUBO results}
\end{tabular}
\label{tab:test_library_and_manual}
\end{table*}


\paragraph{Initial Custom Implementation Results.}
Column \textit{Initial Custom} of Table~\ref{tab:test_library_and_manual} reports the results obtained with our custom implementation.  Although execution times are about 70\% shorter, the precision is significantly degraded:  objective values vary widely, QSAP constraints are not satisfied, and optimal assignments are not reached.

\paragraph{Preliminary Analysis.}
The results suggest that while both implementations follow equivalent theoretical principles, their performances diverge. The custom implementation achieves lower computational cost but produces results 
that are not optimal.  
To understand whether this discrepancy results from circuit design, optimizer settings, quantum primitives, or post-processing, we carried out a systematic comparison between the two implementations. The following section (\S\ref{sec:systematic-investigation-of-the-root-cause}) presents the results of this systematic comparison.

\section{Systematic Investigation of the Root Cause}
\label{sec:systematic-investigation-of-the-root-cause}


To investigate the cause of the discrepancy between the two implementations (i.e., baseline and custom), we compared their equivalent components. 
Because both implementations take a QUBO matrix as input and return a QUBO assignment vector as output, the transformation of the problem into a QUBO and the display of results are excluded from the comparison. 
The following components were analyzed in detail:

\begin{enumerate}[label=\textbf{\arabic*.}, leftmargin=1.2em, itemsep=0pt, topsep=1pt]
\item the quantum circuit (with and without parameters);
\item the parameters of the classical optimizer;
\item the quantum primitives and their configurations;
\item the post-processing applied to the circuit outcomes.
\end{enumerate}

\subsection{Quantum Circuit Equivalence}
Although the two circuits are not identical, their functional behavior is equivalent. 
To confirm this equivalence, we compared both circuits using two complementary methods:
\begin{enumerate}[label=\textbf{\arabic*.}, leftmargin=1.2em, itemsep=0pt, topsep=1pt]
    \item the \textit{state vector}, to ensure that both circuits produce identical final quantum states;
    \item a \textit{heatmap analysis}, displaying the amplitude and phase components of their unitary matrices in complex space.
\end{enumerate}
Both analyses confirmed that the circuits are functionally equivalent and therefore not the source of the accuracy discrepancy.

\subsection{Classical Optimizer Parameters}
Both implementations use the COBYLA optimizer with identical hyperparameters:
\begin{itemize}[leftmargin=1.1em, itemsep=0pt, topsep=1pt]
    \item \textbf{Tolerance:} convergence threshold defining when minimization stops (none specified in our tests);
    \item \textbf{Maximum iterations:} the number of optimizer evaluations of the quantum circuit (1\,000 in our tests);
    \item \textbf{Learning rate:} the step size controlling how far the optimizer moves in parameter space (1 in our tests).
\end{itemize}
Since these settings were identical, the classical optimization process cannot explain the performance gap.

\subsection{Quantum Primitives}
The comparison of primitives revealed one fundamental difference: our implementation explicitly uses two primitives—\texttt{Estimator} and \texttt{Sampler}—while the library relies solely on a \texttt{Sampler}. 
At first glance, this might suggest that the library implicitly defines its own estimator. 
Indeed, internal inspection shows that Qiskit Algorithms calls a private class functioning as an estimator but implemented on top of its sampler primitive. 
Although undocumented, this internal estimator appears optimized for diagonal Hamiltonians, which fits the characteristics of QAOA problems.

Another difference concerns the primitive versions. The library uses the first version of Qiskit primitives, which allows an unlimited number of sampling shots in simulation mode. 
In contrast, our implementation uses the second version of primitives, which limits the sampling budget per execution. 
We set this number to 10\,000—a large but finite value consistent with realistic quantum hardware execution. 

To isolate the impact of primitives, we tested our implementation using the same internal estimator and the first version of the sampler, reproducing the library’s configuration exactly. Despite these changes, the accuracy difference persisted. Therefore, primitives are not the primary source of performance gap.



\subsection{Post-Processing and Sampling Budget: Root Causes of Accuracy Loss}

The accuracy gap between implementations stems from two interrelated factors: post-processing extent and sampling budget. Inspecting the Qiskit Algorithms source code revealed an undocumented detail: all sampled results form a probability distribution, and any bitstring with frequency $\geq 1\times10^{-6}$ is classically evaluated before selecting the lowest-cost assignment, as shown in Listing~\ref{lst:post-processing}.

\begin{lstlisting}[language=Python, caption={Post-processing classical evaluation in the library}, label={lst:post-processing}]
# From qiskit_algorithms (simplified)
def _post_process_samples(self, samples):
    threshold = 1e-6  # Undocumented!
    filtered = {k: v for k, v in samples.items() 
                if v >= threshold}
    costs = {k: evaluate_qubo(k, self.qubo) 
             for k in filtered}
    return min(costs, key=costs.get)
\end{lstlisting}

This threshold's effectiveness depends critically on sampling density. With unlimited sampling budget, the library implementation produces exact probability distributions where approximately 31,434 of 32,768 possible states (96\%) exceed the threshold and undergo classical evaluation. In contrast, our custom implementation with 10,000 shots sampled only 10 assignments ($\approx 0.03\%$). This sparse sampling reduces the effective search space available to post-processing.

Column \textit{Custom-Post-Proc-threshold-corrected} in Table~\ref{tab:test_library_and_manual} confirms this analysis: replicating the library's threshold with 10,000 shots improved results (4/10 optimal solutions, mean objective 660) but remained below the library baseline due to limited state coverage. The comparison demonstrates that post-processing plays a decisive role in solution quality—evaluating more candidate assignments increases accuracy but also extends classical computation time. However, this filtering capability fundamentally depends on adequate sampling: without sufficient shots to populate the probability distribution, even optimal post-processing strategies cannot recover missing high-quality states.

These observations reveal that the sampling budget is the root cause of accuracy differences, with post-processing serving as an amplifier whose effectiveness depends on sampling density. 

\subsection{Validation: Shot Scaling Experiments}

We conducted 30 trials at increasing shot counts to validate our hypothesis. 
As shown in Table~\ref{tab:shot_scaling}, sampling budget proved the dominant factor in QAOA accuracy. 
Objective values improved monotonically from 642 at 10\,000 shots to 563 at 250\,000, converging to the library’s optimal value. 
Sampling coverage expanded proportionally—from 7,593 evaluated states (23\%) to 28,706 (88\%)—compared to the library’s 97\% coverage with exact probabilities. 
The optimal rate increased from 40\% to 97\%, stabilizing once coverage reached roughly 70–80\%, beyond which additional shots yielded minimal gains. 
Computation time grew sublinearly, remaining around 68 s despite a 25× increase in shots, confirming that classical post-processing dominates runtime, whereas the library’s exhaustive evaluation required about 188~s. 

\begin{table*}[htbp]
\caption{Impact of Sampling Budget for Classical Post-Processing on QAOA Results Accuracy (Critical Finding) With 30 executions for Each Shot Number and Post-Processing Threshold Correction}
\small  
\label{tab:shot_scaling}
\centering
\begin{tabular}{ccccccc}
\toprule
\textbf{Shots} & \textbf{Avg Time (s)} & \textbf{Avg Objective} & \textbf{Std Dev} & \textbf{Optimal Rate} & \textbf{States Evaluated} & \textbf{Coverage} \\
\midrule
10000 & 69.12 $\pm$ 5.59 & 642 & 99 & 40\% & 7593 & 23\% \\
25000 & 68.79 $\pm$ 3.92 & 616 & 70 & 43\% & 13936 & 43\% \\
50000 & 71.23 $\pm$ 5.24 & 592 & 35 & 53\% & 19538 & 60\% \\
100000 & 70.44 $\pm$ 5.56 & 574 & 26 & 80\% & 24362 & 74\% \\
150000 & 72.62 $\pm$ 7.32 & 570 & 25 & 87\% & 26710 & 82\% \\
200000 & 66.42 $\pm$ 3.71 & 565 & 16 & 93\% & 27882 & 85\% \\
250000 & 65.96 $\pm$ 3.43 & 563 & 12 & 97\% & 28706 & 88\% \\
$\infty$ (library) & \textbf{188.30 $\pm$ 7.37} & \textbf{561} & \textbf{0} & \textbf{100\%} & \textbf{31649} & \textbf{97\%} \\
\bottomrule
\end{tabular}
\end{table*}

\section{Analysis and Implications}
\label{sec:analysis-and-implications}
The comparative analysis in Section~\ref{sec:systematic-investigation-of-the-root-cause} highlights three critical challenges: clarifying the quantum–classical interface, making implicit parameters explicit, and optimizing post-processing balance. The following sections provide practical recommendations and broader challenges for quantum software engineering.

\subsection{Understanding the Quantum-Classical Interconnection }

\begin{figure}[htbp]
\centerline{\includegraphics[scale=0.3]{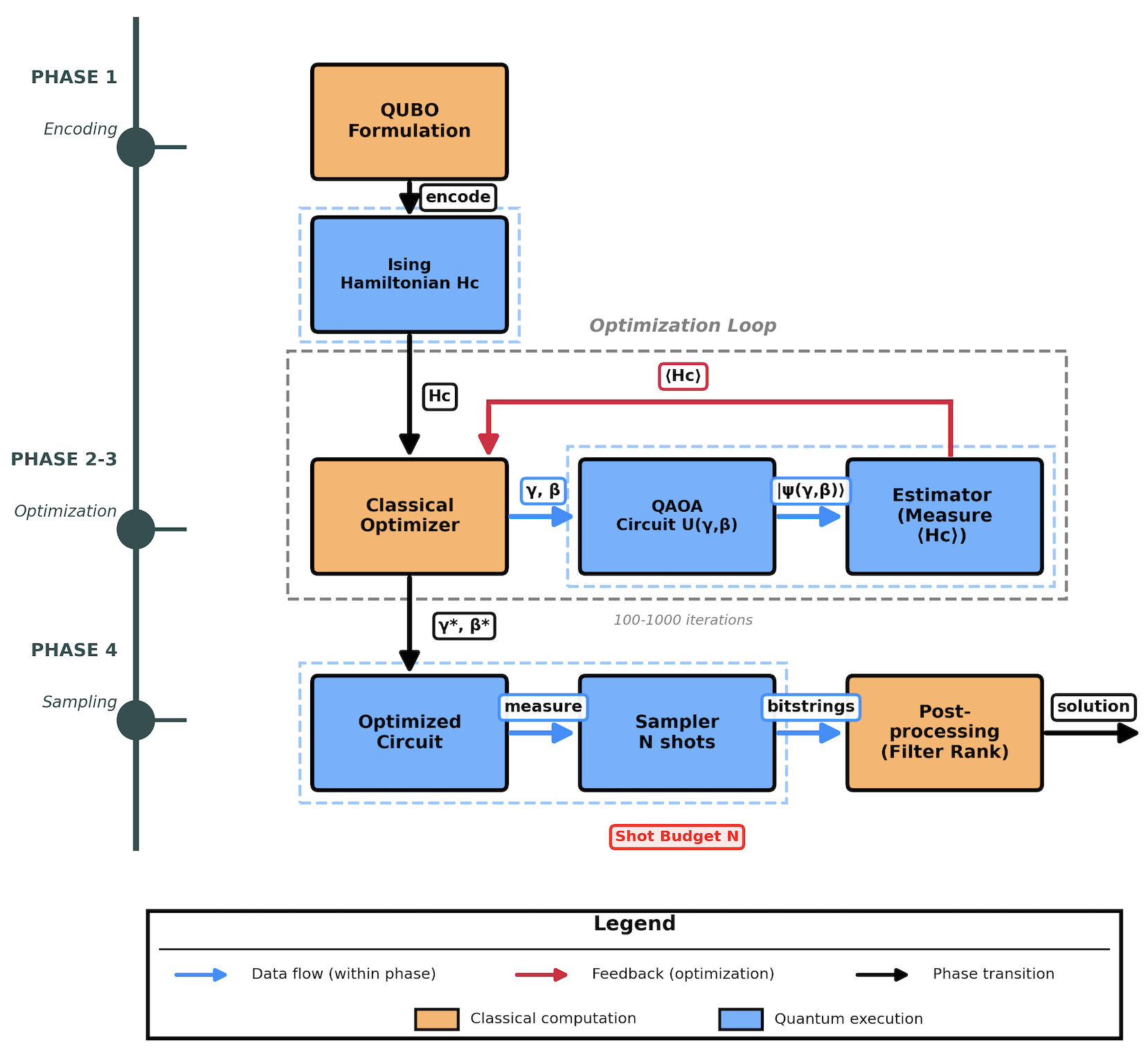}}
\caption{Qiskit QAOA implementation architecture. Classical optimizer interacts with quantum backend through Qiskit primitives: Estimator (Phases 2-3) for measuring $\langle H_C \rangle$ during optimization, and Sampler (Phase 4) for collecting bitstrings with shot budget $N$.  }
\label{fig:hybrid_interface}
\end{figure}

Figure~\ref{fig:hybrid_interface} illustrates the QAOA implementation in Qiskit, revealing how the framework structures quantum-classical interactions across four phases. Phase 1 maps the LAR problem to the Ising Hamiltonian $H_C$ using Qiskit's encoding utilities. Phases 2-3 implement the variational optimization through Qiskit's hybrid interface: a classical optimizer (e.g., COBYLA) provides parameters $(\gamma, \beta)$ to the quantum circuit, which is executed via the Estimator primitive to measure $\langle H_C \rangle$. This expectation value feeds back to the optimizer through Qiskit's callback mechanism, iterating 100-1000 times until convergence. Phase 4 leverages the Sampler primitive to execute the optimized circuit with a specified shot budget $N$, producing measurement bitstrings that are processed by Qiskit's classical post-processing layer to filter and rank solutions. The quantum circuit serves as a probabilistic filter that directs classical computation toward meaningful regions of the search space.

\subsection{Understanding Implicit Parameters}

The sampling budget dependency was hidden by several factors. First, shot configuration in Qiskit is not required. Users could call the \texttt{run()} method of the sampler without specifying shots, never realizing it's a tunable parameter affecting accuracy. Second, statevector simulators can return exact probabilities, making shots seem optional. Third, neither Qiskit Algorithms nor v2 primitives documentation emphasizes the accuracy-shots relationship. The v2 default (1\,024) seems arbitrary rather than principled.

\subsection{Balancing Post-Processing Effort}
Post-processing strongly affects both accuracy and runtime. 
In our experiments, expanding classical evaluation from 23\% to 97\% of possible assignments improved accuracy by over two orders of magnitude but increased computation time proportionally. 
This trade-off depends on the problem’s objective landscape rather than a fixed optimal level. 
As shown by Zhou \textit{et al.}~\cite{zhou2020quantum}, sampling and measurement strategies largely determine hybrid algorithm accuracy. 
Adaptive post-processing—adjusting the number of classically evaluated candidates based on convergence—offers a promising direction. 
Post-processing must be tuned to both problem structure and sampling density to balance accuracy and efficiency.

\subsection{Practical Recommendations}

\subsubsection{For Algorithm Developers}

\textbf{Recommendation 1: Experiment with different Sampling Budgets.} 
The sampling budget determines accuracy, reproducibility, and comparability of hybrid quantum algorithms. Developers should explicitly document shot counts, report statistical measures (mean and standard deviation) over multiple runs rather than single outcomes, and compute coverage metrics indicating the proportion of computational basis states sampled. Prior analyses~\cite{cerezo2021variational} show sampling requirements grow sub-exponentially with system size.

\textbf{Recommendation 2: Validate End-to-End Pipelines.}  
Circuit equivalence alone does not ensure the same behavior. It is important to check the full pipeline: the unitary matrices, optimizer settings and learning rates, sampling budgets, post-processing thresholds, and the random seeds used for reproducibility.

\subsubsection{For Framework Developers}

\textbf{Recommendation 3: Make Sampling Budget Explicit and Required.}  
Frameworks should require explicit sampling budget specification. Hidden defaults risk inconsistent results, especially when migrating between simulators and hardware where statistical noise differs. Explicit specification ensures transparency and prevents silent failures. 
APIs should at least enforce providing the sampling budget, and ideally support adaptive computation of the sampling budget by enforcing a coverage. This is exemplified by Listing~\ref{lst:explicit_shots}. 

\begin{lstlisting}[language=Python, caption={Explicit and adaptive shot specification in Qiskit-style APIs.}, label={lst:explicit_shots}]
# Bad (implicit default)
# 1024 shots by default in Qiskit
sampler.run(circuit)
# Good (explicit specification)
sampler.run(circuit, shots=10000)
# Better (adaptive coverage target)
sampler.run(circuit, coverage_target=0.95)
# -> Automatically determines the sampling budget required
\end{lstlisting}

\textbf{Recommendation 4: Document Behavioral Changes in Version Migrations.}  
Migration documentation should describe changes affecting performance or reproducibility, not just syntax. This includes behavioral differences (e.g., “v2 primitives require explicit shots”), performance implications (e.g., “1024 or even 10k shots may provide insufficient coverage”), and equivalence verification. Clear guidance enables result reproduction across versions and explains deviations from modified execution semantics.

\subsection{Broader Challenges for Quantum Software Engineering}

This case study reveals several quantum software engineering challenges:

\begin{enumerate}[label=\textbf{\arabic*.}, leftmargin=1.2em, itemsep=0pt, topsep=1pt]
\item \textbf{Abstraction Leakage:} Quantum–classical interconnections may require parameters that cause orders-of-magnitude performance differences. Leaky abstractions shift developer attention to sampling and noise management. Quantum software thus needs “transparent abstractions” exposing critical parameters.

\item \textbf{Testing Challenge:} Traditional testing checks correctness, but hybrid algorithms can be functionally correct yet useless due to parameter misconfiguration. New testing paradigms must assess performance, not just correctness.

\item \textbf{Framework Evolution Risks:} Quantum frameworks evolve faster than classical stacks, creating maintenance burden and risking production failures. The field needs versioning and deprecation policies balancing innovation with stability.

\item \textbf{Documentation Inadequacy:} Quantum algorithms are described mathematically (circuits, Hamiltonians) but rarely operationally (shot configuration, tolerance, failure handling). Bridging this gap requires collaboration between quantum researchers and software engineers.
\end{enumerate}
\section{Limitations}
\label{sec:limitations}

This study has limitations affecting generalizability. First, experiments used a single 15-qubit LAR instance. While preliminary tests on 10- and 20-qubit problems show consistent scaling between sampling density and accuracy, broader validation is required across problem classes (MaxCut, TSP, portfolio optimization), circuit depths ($p = 1$–$20$), and optimizers (SPSA, Adam, L-BFGS-B). Second, noiseless statevector simulation was used; real hardware introduces gate/measurement errors, decoherence, and crosstalk, which alter quantum-classical balance and reduce reproducibility. Third, only COBYLA was tested; other optimizers may show different sampling budget sensitivities, with gradient-based methods like SPSA potentially requiring fewer shots due to stochastic gradients. Future work should systematically benchmark on noisy hardware across diverse problem instances.

\section{Related Work}
\label{sec:related}

Zhao~\cite{zhao2020quantum} and Miranskyy et al.~\cite{miranskyy2021quantum} identified quantum software engineering as an emerging discipline with unique challenges. Ali et al.~\cite{ali2022testing} surveyed testing techniques for quantum programs, noting the difficulty of validating probabilistic outputs—our case study exemplifies this: correct circuits producing incorrect results due to misconfigured sampling. Weder et al.~\cite{weder2021hybrid} emphasized explicit quantum-classical interfaces in hybrid workflows; our findings confirm sampling is a critical interface where configuration determines system behavior. Leymann and Barzen~\cite{leymann2020bitter} discuss parameterization and measurement patterns; we extend this by showing sampling patterns (shot counts, timing) deserve equal attention as first-class design concerns.
Classical framework migration is well-studied through API refactoring patterns~\cite{dig2006api} and deprecation practices~\cite{hora2015deprecation}. Spinellis~\cite{spinellis2018software} found interface stability crucial for ecosystem health. However, quantum framework migration remains unexplored. The rapid evolution of quantum software—Qiskit's major versions every 1-2 years versus decade-long classical stability—creates tension between innovation and stability, suggesting the need for quantum-specific migration patterns and tooling.

\section{Conclusion}
\label{sec:conclusion}

Migrating quantum algorithms across framework versions requires validating the entire computational pipeline, not only circuit translation. Our conversion of QAOA from Qiskit 1.x to 2.x showed that accuracy depends strongly on the sampling budget—the number of results from circuit executions with the best parameters. A 10 000-shot limit in v2 primitives produced only 23\% state coverage for classical evaluation and two orders of magnitude worse accuracy than the library baseline. After confirming circuit and optimizer equivalence, increasing the sampling budget to 100 000 greatly improved accuracy, showing that sampling density—not algorithm design—explains the discrepancy. Post-processing plays an essential role in accuracy; for optimization tasks, good results are impossible without evaluating a sufficient number of samples. The post-processing threshold may be adjusted depending on the problem, especially for complex cases where classical evaluation is expensive. Future work will assess performance on real quantum hardware and explore adaptive sampling strategies that adjust shot counts dynamically during optimization.

\begin{acks} 
This work was supported by Mitacs and Pinq2 through the Accelerate Program. 
\end{acks}

\bibliographystyle{ACM-Reference-Format}
\bibliography{references}

\end{document}